\documentclass[sigconf]{acmart}
\usepackage{chato-notes}

\usepackage{amsfonts}
\usepackage{amsmath}
\usepackage{nccmath}
\usepackage{graphicx}

\usepackage{dsfont}
\usepackage{booktabs}
\usepackage{multirow}
\usepackage{tabularx}

\usepackage{cleveref}
\usepackage{hyperref}
\hypersetup{colorlinks=true,citecolor=brown}
\usepackage{url}
\usepackage{xcolor}

\copyrightyear{2024}
\acmYear{2024}
\setcopyright{acmlicensed}\acmConference[WWW '24 Companion]{Companion Proceedings of the ACM Web Conference 2024}{May 13--17, 2024}{Singapore, Singapore}
\acmBooktitle{Companion Proceedings of the ACM Web Conference 2024 (WWW '24 Companion), May 13--17, 2024, Singapore, Singapore}
\acmDOI{10.1145/3589335.3651980}
\acmISBN{979-8-4007-0172-6/24/05}

\acmSubmissionID{107}

\definecolor{mygray}{rgb}{0.75, 0.75, 0.75}

\newcommand{\node}{c} 

\newcommand{\feat}{\mathbf{x}}

\newcommand{\user}{u}

\author{
Andreas Damianou, Francesco Fabbri, Paul Gigioli, Marco De Nadai, \\
Alice Wang, Enrico Palumbo, Mounia Lalmas}
\def \authors{Andreas Damianou, Francesco Fabbri, Paul Gigioli, Marco De Nadai, Alice Wang, Enrico Palumbo, Mounia Lalmas}
\affiliation{%
\institution{Spotify
\country{}
}
}

\begin{document}

\title{Towards Graph Foundation Models for Personalization}


\renewcommand{\shortauthors}{Damianou \emph{et al.}}

\begin{abstract}
In the realm of personalization, integrating diverse information sources such as consumption signals and content-based representations is becoming increasingly critical to build state-of-the-art solutions. 
In this regard, two of the biggest trends in research around this subject are Graph Neural Networks (GNNs) and Foundation Models (FMs). 
While GNNs emerged as a popular solution in industry for powering personalization at scale, FMs have only recently caught attention for their promising performance in personalization tasks like ranking and retrieval.
In this paper, we present a graph-based foundation modeling approach tailored to personalization. 
Central to this approach is a Heterogeneous GNN (HGNN) designed to capture multi-hop content and consumption relationships across a range of recommendable item types. To ensure the generality required from a Foundation Model, we employ a Large Language Model (LLM) text-based featurization of nodes that accommodates all item types, and construct the graph using co-interaction signals, which inherently transcend content specificity. To facilitate practical generalization, we further couple the HGNN with an adaptation mechanism based on a two-tower (2T) architecture, which also operates agnostically to content type. This multi-stage approach ensures high scalability; while the HGNN produces general purpose embeddings, the 2T component models in a continuous space the sheer size of user-item interaction data. Our comprehensive approach has been rigorously tested and proven effective in delivering recommendations across a diverse array of products within a real-world, industrial audio streaming platform.
\end{abstract}



\maketitle

\section{Introduction}
\label{sec:intro}
Foundation models have emerged as capable approaches in recent years, resulting in unprecedented success across a broad spectrum of applications, ranging from Natural Language Processing (NLP) to Computer Vision and Audio Processing~\cite{yuan2021florence, xiao2023florence, lu2023unified,brown2020language, gardner2023llark}. 
A Foundation Model (FM) is a large-scale pre-trained neural network architecture, typically based on a Large Language Model (LLM), designed to serve as a base or foundation for various downstream tasks~\cite{bommasani2021opportunities}. The ability to perform on a wide range of tasks is largely attributed to their pre-training on vast amounts of data. To leverage the capabilities of such a model, fine-tuning is typically applied to further enhance its performance on specific tasks or domains. 
Only lately in search and recommendation, or more broadly in personalization, LLMs have gained growing attention, largely attributable to their ability to map user preferences into natural language. Recent research underscores the effectiveness of employing fine-tuning strategies or leveraging few-shot-learning with LLMs, both of which have been shown to yield competitive results in personalization tasks~\cite{zhang2023collm, lyu2023llm, geng2022recommendation}.
On the other hand, even if promising, these approaches become challenging when the recommendations are performed at scale because they struggle to adapt quickly to catalog changes (e.g. when new items are introduced or when user preferences change).

On the other hand, graph-based learning models, specifically Graph Neural Networks (GNNs), have emerged as a powerful technology for recommendation systems at scale, becoming a core functionality on different online and social platforms~\cite{ying2018graph, el2022twhin, halcrow2020grale, xie2023graph}. 
Moreover, only lately, GNNs have been showing relevant gains also for enabling discovery without loss in accuracy~\cite{denadai2024}.
Their success is often attributed to their capacity to explicitly model long-range and heterogeneous relationships, content semantics, and content features simultaneously. 
Furthermore, GNNs showcase strong inductive capabilities, meaning they can generalize knowledge from training data to unseen entities, a trait that is crucial for tackling complex and evolving datasets, essential for applications such as recommendation systems and social networks. 

Graph foundation models (GFMs) is a novel and promising direction that aims to bring the foundation modeling capabilities to the domain of graph learning. However, there is still no clear consensus in the community regarding a ``definition'' of a GFM, although such a model would be expected to at least be able to generalize and adapt across tasks~\cite{liu2023towards}. Currently, there are limited published GFM approaches, and their versatility does not typically reach that of FMs for NLP or Computer Vision. This can largely be attributed to the fact that a graph topology is a more complicated and arbitrary structure compared to the ubiquitous sequential (e.g.~textual or visual) representation leveraged by LLMs. To make progress, various more specialized notions of generalization and transferability have been explored: \citep{galkin2023towards} propose a task-specific FM model for KG completion that inductively generalizes to unseen graphs; Huang et al.~\cite{huang2023prodigy} build in-context learning mechanisms over graphs. 

In this paper we focus our attention on GFMs applied to the broad domain of personalization. ``Personalization'' refers to the process of tailoring item recommendations or search results to individual users. For example, a recommender system can suggest an item $\node$ to a user $\user$ based on the item's features $\feat_\node$ that capture its content characteristics and the user's features, $\feat_\user$, that capture the user's characteristics, past behavior etc. As a GFM, our approach possesses the properties of generalization and adaptability across tasks; however, these properties are specialized for the general domain of personalization rather than applicable to any possible task and graph (e.g.~knowledge graphs, biology, molecular graph generation). This specialization allows us to make design choices that render the GFM approach practical and scalable. Given the GFM taxonomy proposed in~\citep{mao2024graph}, our approach would fall in the category of ``domain-specific GFM''. The ingredients that compose our approach are a heterogeneous GNN (HGNN) with LLM-featurized node features and an adaptation (across tasks) mechanism based on a two-tower (2T) user/item model that, contrary to common 2T architectures, operates agnostically to item type. 

As a running example, which constitutes the motivation of our work and corresponds to the experiments shown later on, we will consider a real-world industrial audio streaming platform with a massive catalog of users and content. The users can interact with items of different types, such as podcasts or audiobooks. 

\textbf{Our Contribution.} To the best of our knowledge, we propose the first GFM that acts as foundation specifically for solving various personalization tasks such as recommendations of diverse types of items and flavors such as ``similar to your item X'' or ``based on your history'', as well as judging suitability of search results. Before adapting to tasks, such a GFM can be pre-trained on an entire catalog, which includes information about items' content and past user-interactions. By combining (inductive) GNNs and LLMs we ensure that the overall foundation model can generalize and extrapolate well even if it is only trained once -- subsequent updates can be relatively infrequent. The 2T component enables scalability and adaptation of the foundational generalized representations. The benefit of such an approach is that it unifies representation learning across various tasks, it enables information sharing, improves the quality of learned representations, simplifying production pipelines. This comes in contrast to traditional personalization approaches that develop siloed solutions for different item types or tasks. Finally, our work is motivated by and tested on a real-world industrial application, and thus it constitutes a valuable case-study on the design of scalable and industry-appropriate GFMs.

\section{GFM for Personalization}

We describe our proposed two-step architecture, comprising a HGNN/LLM foundation core and a two-tower (2T) adaptation mechanism. The two components correspond to the introduced notions of a ``static'' and a ``dynamic'' layer. After describing the details of the approach, we explain how it has been applied on our motivating scenario of personalization in an audio streaming platform.

\begin{figure*}[t]
     \includegraphics[width=0.9\textwidth]{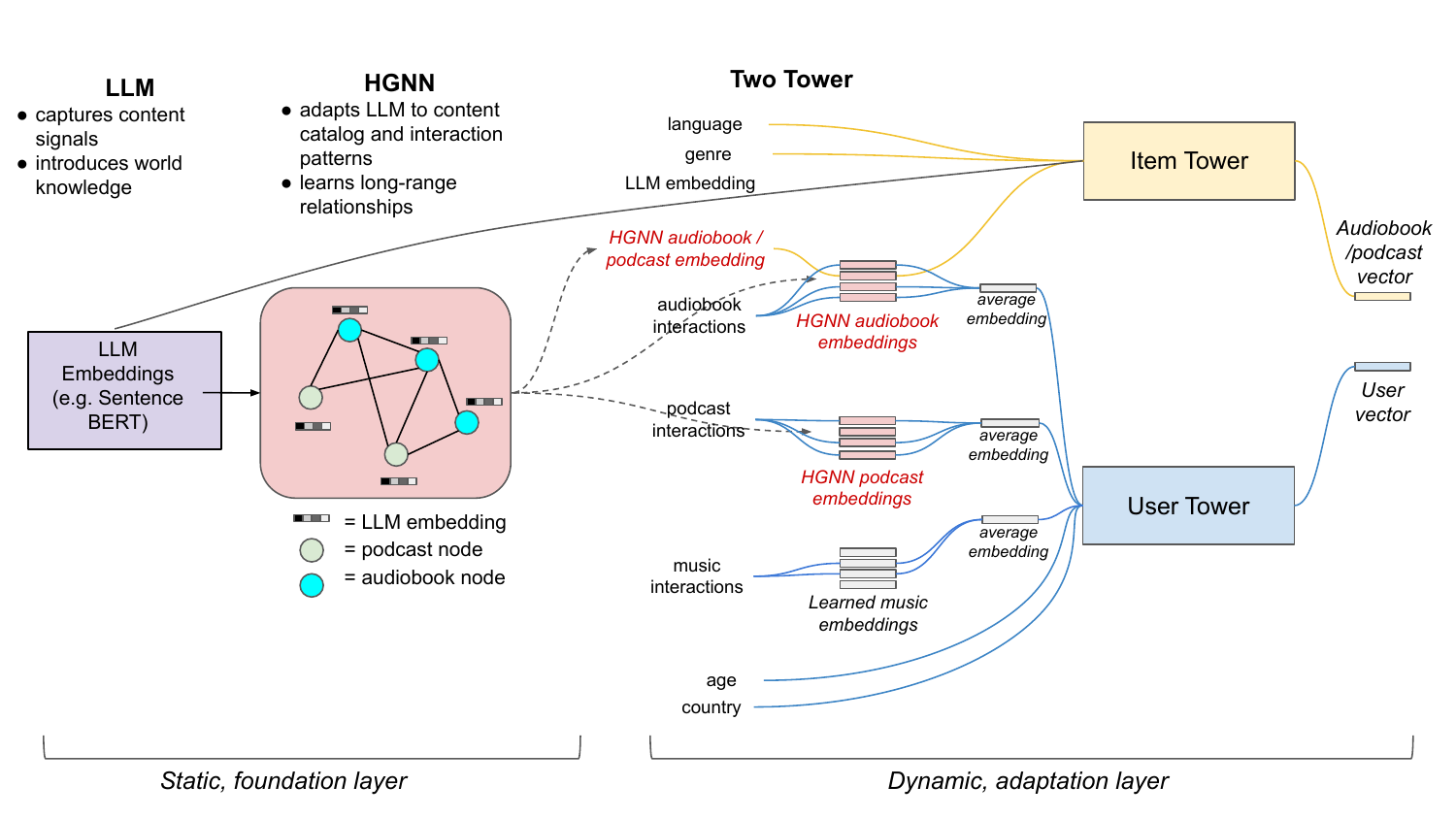}
     \caption{Our GFM-based approach to personalization, instantiated for the domain of audiobook/podcast recommendations in a real-world online streaming platform. The LLM embeddings are passed as node features for the HGNN as well as item features in the 2T. The corresponding HGNN embedding for a given item is passed to the item tower. For a given user, we take the average of the HGNN embeddings corresponding to all items that the user has interacted with. Notice how the HGNN embeddings are, thus, shared between the two towers. With this approach, the resulting user and item embeddings (irrespective of item type) all lie in the same vector space.}
     \label{fig:overall}
 \end{figure*}

\textbf{Static (foundation) layer.} This is an HGNN trained on an item-item graph, where items are connected if they have been co-interacted by the same user. A variety of node and edge types is considered. Each node is associated with text embedding node-features coming from a general LLM applied on the item's  description. The LLM embeddings allow any type of node to be uniformly represented in the graph. Crucially, interaction signals are part of the graph but, still, users are not represented as nodes. This allows us to (a) keep the HGNN representation generic and static, meaning that it only needs to be updated infrequently, and (b) scale to very large databases, deferring the task-specific user representation to the downstream fine-tuning adaptation. This is the foundational component of our approach; it combines content with interaction signals and is generic since it can yield general purpose embeddings. 

\textbf{Dynamic (adaptation) layer.} The foundation static layer learns embeddings that represent each item node in the graph, which are then passed as input to a 2T (item/user) model. The 2T model constitutes the mechanism by which foundational representations are adapted for a variety of downstream tasks. Because it is lightweight, it allows for adaptation to user preferences at a large scale, reflected in a continuous stream of interaction data. Hence, this component is referred to as the dynamic layer of the overall architecture and is trained often. Architecturally, a 2T model~\cite{yi2019sampling} consists of an item tower that encodes content level information and a user tower that encodes user demographics as well as item interactions. Item interactions can be represented as learned embeddings where each item identifier is mapped to a row in an embeddings matrix. This embeddings matrix is also used by the item tower, which creates a shared embeddings layer between the item and user tower. Traditionally, different item towers are considered for each type of item. Here, we introduce a single, type agnostic item tower, which heavily leverages the HGNN and LLM embeddings. 

To exemplify the usage of our HGNN-based foundation model, we return to our motivating challenge of driving personalization in a real-world, industrial audio streaming platform.\footnote{This architecture can be applied to other similar domains.} The approach is illustrated in Figure~\ref{fig:overall}. Here, the platform's catalog and interaction data are represented as a graph that captures the top-level item type information, such as podcast-podcast, audiobook-podcast and audiobook-audiobook relationships. Edges are added between nodes whenever at least one user interacts with both items.  Each node is associated with a node type, and multiple types of relations between nodes are possible. To represent content, we incorporate node features through LLM Sentence BERT embeddings. A HGNN is trained on the graph, distilling as HGNN embeddings the interaction signals (used to form edges) with item content signals (node features), leveraging multi-hop relationships found in the graph. Our implementation is based on GraphSAGE~\cite{hamilton2017inductive}, and we refer the reader to the paper for more details. The model is trained with a self-supervised link-prediction loss, similar to Ying \emph{et al.}~\cite{ying2018graph}, to refine node representations, ensuring they not only reflect their self-presentation independently of the graph structure but also align with the representations of all connected nodes within the~graph. 

The 2T model comprises two feed-forward deep neural networks—one for users and the other for items. This model utilizes the embeddings learned from the HGNN, minimal item metadata, and user metadata within the user tower. Within the user tower we also represent user interests on audiobooks and shows by averaging the HGNN vectors that the user has interacted with in the last 90 days. We also add music interests as the average of music vectors learned by the company owning the data. We aim to train the model in such a way that user vectors closely match the content-agnostic item vectors and enhance recommendation accuracy.

Notice that the item tower of the 2T component of our approach treats audiobooks and podcasts agnostically, since it just expects a HGNN and a LLM embedding of a certain dimension. Further, lower-level item types, such as podcast episodes, can be used in a \textit{zero-shot} fashion: an episode inherits the HGNN embedding of its corresponding show. Similarly, inference for new items in the catalog is possible thanks to the inductive capabilities of the HGNN. 

Overall, this static-dynamic architecture allows us to combine the robust representation power of the 
HGNN with the scalable and adaptable characteristics of the two tower model.  Moreover, the unified representation of all content types within the 2T model allows us to (a) have all item embeddings in the same vector space, (b) treat the 2T model as a lightweight FM task-adaptation mechanism, (c) leverage overlapping information shared by the content types, and finally (d) mitigate bias (e.g. popularity) by exporting a lot of the content representation learning to the foundation HGNN model, in effect de-coupling content representation learning from user representation learning.

\section{Related work} 

While our work is, to our knowledge, the first to discuss GFMs specifically for the domain of personalization, notable prior work exists in the general area of GFMs. 
Recent efforts have introduced the definition of GFMs~\cite{liu2023towards, mao2024graph}, with the first also proposing a taxonomy based on the type of technology powering the architecture. In this work we propose a model powered by a static layer combining a GNN and an LLM, showing how a synergy between the two architectures can help to combine content and consumption patterns. In a similar fashion, Xie et al.~\cite{xie2023graph} in their work show how a graph-aware Language Model framework can help to improve performances on different downstream tasks on large scale industry data. The static layer proposed in our work can be seen as analogous to the pre-training architecture proposed by Xie et al. It is worth noting that there are multiple previous efforts combining graphs with LLMs, but the focus has been less on creating a foundation model. Notable works include joint training of GNNs and LLMs at Amazon \cite{ioannidis2022efficient} and combining KGs with LMs \cite{zhang2021greaselm, yasunaga2022deep}.

Galkin et al.~\citep{galkin2023towards} propose ULTRA, a FM task-specific model for knowledge graph completion. Our paper proposes a GFM that is domain-specific, as we aim to perform different downstream tasks (e.g. ranking and recommendation) across multiple item types for one domain (personalization).

Our work focuses on the importance of relational data, including consumption-based signals, to unlock content semantic understanding and building the graph foundation model for personalization. Fey et al.~\cite{fey2023relational} recently highlighted how classical machine learning methods struggle to learn from interconnected data sources, requiring manual crafting and feature engineering.  They introduce the paradigm of Relational Deep Learning (RDL), which offers a streamlined and improved solution without the need for extensive feature engineering. In this work we bring further evidence of the importance of using GNN-based solutions to aggregate different sources of information, without requiring complex feature engineering.

Graph-based methods for search and recommendation at scale have demonstrated their effectiveness not only in optimizing accuracy but also in enhancing the impact on long-tail items. Recent work by Palumbo et al.~\cite{palumbo2023graph} shows how graph learning methods effectively diversify without compromising accuracy. Similarly, in the work by De Nadai et al.~\cite{denadai2024}, a combination of a heterogeneous GNN with a standard 2T model is designed to power recommendations for a new product on an online platform. The GNN helps leveraging consumption patterns of a quite consolidated and mature product, to boost start rates for a new one.

\section{Experiments and Results} 

In this section, we present experimental results supporting the usefulness and practicality of the proposed FM for personalization. As a reminder, the 2T component is agnostic to item types and the HGNN component is motivated by the generality of a GFM. In this way, all item types can be represented in a \textit{unified} vector space. Henceforth, we will refer to this approach as the \textit{Unified 2T model}. 

For our experiments we consider as a dataset a sample of 10M users, 3.5M podcasts and 250K audiobooks from an online audio streaming platform. We use 90 days of data to train the model and a hold-out dataset (for evaluation purposes) comprising all the audiobook and podcast streams of users in the last 14 days. We evaluate the performance of our recommendation task with Hit-Rate@K (HR@K), where~$K = 10$. 

Our investigation first explores the generalization capability of the Unified model. For this purpose, we trained a 2T baseline model focused solely on audiobook recommendations and compared it with our Unified model, which is trained on both audiobook and podcast recommendations. Despite the diverse content types handled, both models employ identical input features and comparable hyperparameters. As shown in Table~\ref{tab:audiobooks}, the Unified model outperforms the content-specific model in performance, highlighting its superior generalization ability even when trained on multiple content types simultaneously. Note that the Unified model uses only LLM embeddings without any specific adjustments or adaptations between content types.

Then, we explore the impact of removing GNN content representation from the Unified model. In this baseline, we learn content vectors from scratch during training. Table~\ref{tab:gnn} highlights the critical role of the HGNN foundational representation, which significantly improves the recommendation of both podcasts and audiobooks. Note that Table~\ref{tab:audiobooks} and Table~\ref{tab:gnn} cannot be compared to each other as the models have been trained on different days.

Finally, we challenge the previously discussed differentiation between the static and dynamic layers of the model. We compare our model to a variant in which both the 2T and HGNN components are trained daily. Similarly to findings from seminal works in NLP and Computer Vision regarding FMs \cite{bommasani2021opportunities}, the results presented in Table~\ref{tab:stability} confirm that the HGNN foundation representation remains stable over time and can be effectively utilized in the Unified 2T model on a daily basis without the need for frequent retraining.

\begin{table}[t]
	\caption{Audiobook specific model vs. unified model}
    \renewcommand{\arraystretch}{1}
	\centering
	\begin{tabularx}{\columnwidth}{@{}Xrrr rrr@{}}
	\toprule
	\textbf{Model} & \textbf{HR@10} \\ \midrule
        Audiobooks 2T & 0.271 \\ 
        Unified 2T & 0.316 \\
	\bottomrule
	\end{tabularx}
	\label{tab:audiobooks}
\end{table}

\begin{table}[t]
	\caption{Unified model without GNN embeddings}
    \renewcommand{\arraystretch}{1}
	\centering
	\begin{tabularx}{\columnwidth}{@{}Xrr@{}}
	\toprule
	\multirow{2}{*}{\textbf{Model}} & \multicolumn{2}{c}{\textbf{HR@10}} \\ \cmidrule(lr){2-3} 
	& Podcasts & Audiobooks \\
	\midrule
        Unified 2T w/o GNN & 0.159  & 0.329 \\
        Unified 2T & 0.165  & 0.343 \\
	\bottomrule
	\end{tabularx}
	\label{tab:gnn}
\end{table}

\begin{table}[t]
	\caption{Unified model without daily retraining}
    \renewcommand{\arraystretch}{1}
	\centering
	\begin{tabularx}{\columnwidth}{@{}Xrr@{}}
	\toprule
	\multirow{2}{*}{\textbf{Model}} & \multicolumn{2}{c}{\textbf{HR@10}} \\ \cmidrule(lr){2-3} 
	& Podcasts & Audiobooks \\
 \midrule
        Unified 2T & 0.151  &  0.284 \\
        Unified 2T w/o retraining & 0.152  & 0.284  \\
	\bottomrule
	\end{tabularx}
	\label{tab:stability}
\end{table}

Overall, our experiments confirm the effectiveness of our FM for personalization. This model demonstrates that it can generalize well across various content types. It consists of both static and dynamic layers, which are both crucial for accurately representing content and users. Notably, the static layer can be pre-trained through self-supervision at infrequent intervals, confirming its similarity with seminal literature in the FM field.

\section{Conclusions and future work}
In this study we proposed, to the best of our knowledge, the first GFM-based approach tailored to personalization at scale. Our blueprint introduces a novel differentiation between static and dynamic layers, enabling the exploitation of consumption and content patterns at scale. This approach yields high-quality representations, facilitating seamless execution of diverse downstream personalization tasks. While this work marks an initial step towards consolidating GFMs for personalization, there are several suggestions for improvement that open different lines of future work. First, the LLM and GNNs in the static layer are trained independently. However, enhancing synergy between these architectures could involve designing an end-to-end (e2e) solution where the final layers of the LLM and the GNN weights are jointly trained. 
In addition, the e2e training idea can be extended to also include the dynamic layer, with the objective of finding a suitable solution to scale not only content, but also user representation.
Evaluation-wise, we focused on recommendation tasks with multiple content-types. As future work, we plan to collect further insights by extending our experimentation to additionally include search-related tasks, such as query suggestions. This can be achieved with few modification of the existing framework, thanks to its generality.

\bibliographystyle{abbrv}
\bibliography{bibliography}
\end{document}